\documentstyle[preprint,eqsecnum,prd,aps,epsfig]{revtex}
\begin{document}
\draft
\preprint{
\vbox{
\halign{&##\hfil\cr
	& MADPH-95-922 \cr
	& UTTG-13-95 \cr
	& hep-ph/9512320 \cr}}
}
\title{Hadronic $\psi$ production calculated
in the NRQCD factorization formalism}
\author{Sean Fleming}
\address{Department of Physics, University of Wisconsin, Madison, WI 53706}
\author{Ivan Maksymyk}
\address{Theory Group, Department of Physics, University of Texas,
         Austin, TX 78712}

\maketitle
\begin{abstract}
%
%
The NRQCD factorization formalism of Bodwin, Braaten, and Lepage
prescribes how to write quarkonium production rates
as a sum of products of short-distance
coefficients times non-perturbative long-distance
NRQCD matrix elements.
%
%
We present, in the true spirit of
the factorization formalism, a detailed calculation of the inclusive
cross section for hadronic $\psi$ production.
%
%
We find that in addition to the well known
{\it color-singlet} production mechanisms,
there are equally important
mechanisms
in which the $c\bar{c}$
pair that forms the $\psi$ is initially produced in a {\it
color-octet} state, in
either a ${}^3S_1$, ${}^1S_0$, ${}^3P_0$ or ${}^3P_2$ angular-momentum
configuration.
%
%
In our presentation,
we emphasize the ``matching'' procedure, which
allows us to determine the short-distance coefficients
appearing in the factorization formula.
%
%
We also point out how
one may systematically include relativistic
corrections in these calculations.

\end{abstract}

\vfill \eject

\narrowtext
\section{Introduction}

Bodwin, Braaten, and Lepage (BBL) have devised a rigorous
factorization formalism that places the calculation of inclusive
quarkonium annihilation and production on a solid theoretical
foundation~\cite{BBL}.  Their approach is based on nonrelativistic quantum
chromodynamics (NRQCD)~\cite{cl}.  This is an effective field theory
involving an expansion in derivatives,
constructed so as to be equivalent to QCD for
non-relativistic heavy quark-anti-quark scattering, to any
desired order
in the relative three-momentum
of the heavy quarks.
In Ref.~\cite{BBL}, it is shown that the production rate for
a quarkonium bound state $H$ can be written using a
factorization formula, which is a sum of products, having the
form
%
%
%
$$
\sigma(H) = \sum_n \frac{F_n}{m_Q^{d_n - 4}}
\langle 0| {\cal O}^H_n | 0 \rangle \; .
\nonumber
$$
The ``short-distance" coefficients $F_n$ are obtainable
in perturbation theory as a series in
$\alpha_s(m_Q)$, and can be determined through a calculation
of the production rate of an on-shell $Q\bar{Q}$ pair.  As to
the NRQCD matrix elements
$\langle 0| {\cal O}^H_n | 0 \rangle$,
they encode analytically uncalculable ``long-distance" effects
such as the hadronization of a $Q\bar{Q}$ pair into
bound quarkonium.  The index $n$ labels color and angular-momentum
quantum numbers, as well as the order in the NRQCD momentum expansion.
In contrast to previous approaches, the BBL formalism establishes
a framework within which it is possible to properly handle soft
gluon effects and systematically incorporate relativistic
corrections.

The size of any NRQCD matrix element can be estimated as some
power of the small parameter $v$,
the typical velocity of the heavy quarks in the bound
state.  (For charmonium, $v^2\approx 0.3$.)
Combining $v$-scaling estimates with knowledge
of the size of the $F_n$ (for which the order in
$\alpha_s(m_Q)$ is easily determined), one can deduce the
relative importance of the various terms in the factorization
formula. Keeping only those terms warranted by
experimental precision, one can cast each observable
as a sum involving only a small number of NRQCD matrix elements.
In this way, the important NRQCD matrix elements can
be empirically determined by fitting a set of factorization
formul\ae\ to a body of experimental data.

In this paper we carefully present an example of a calculation in the
factorization formalism.
The quantity that we choose to calculate is the
inclusive hadronic $\psi$ production cross
section, since it is needed for a complete
theoretical description of $\psi$ production at fixed target
experiments.
Using $v$-scaling rules alluded to above,
we find that in addition to
previously calculated color-singlet contributions, there exist equally
important contributions involving the production
of $c\bar{c}$ pairs in a color-octet state.
The color-octet
processes which turn out to be important
in hadronic $\psi$ production
are those in which
the heavy quark pair is produced
in either a ${}^3S_1$, ${}^1S_0$, ${}^3P_0$ or ${}^3P_2$ angular-momentum
state.

Taking the octet-mechanism contributions to hadronic
$\psi$ production as an illustrative
example, we present a systematic procedure for determining the short-distance
coefficients $F_n$
appearing in the factorization formula.
This procedure involves the calculation of
rates for the production of $Q\bar{Q}$ pairs with specific
color and angular-momentum quantum numbers.  These production rates
are expressed as an integral over ${\bf q}$, the relative three-momentum
of the $Q$ and $\bar{Q}$ in the $Q\bar{Q}$ rest frame,
with the integrand being a Taylor expansion in ${\bf q}^2/m_c^2$.
The production rates
are calculated both in full QCD and in NRQCD, and then the results
are ``matched,''
allowing a determination of the $F_n$.

The Feynman diagrams for the leading color-singlet
hadronic $\psi$ production
subprocesses
are shown in Figure 1.
However, we do not calculate these leading color-singlet
contributions since
they have already been treated using the color-singlet-wavefunction
model~\cite{sch}.  For calculations at leading order in
$v^2$, the results of this latter approach can be readily
transformed into the factorization formalism language.

\section{NRQCD factorization approach}

\subsection{Factorization Formula}

As was stated in the introduction,
the NRQCD factorization formalism prescribes a means of expressing
the inclusive production rate of a heavy quarkonium meson $H$
using a factorization formula, which is a sum of products of the form
\begin{equation}
\sigma(A + B \to H + X) = \sum_n \; \frac{F_n}{m_Q^{d_n - 4}} \;
\langle 0|{\cal O}^{H}_n |0  \rangle \; .
\label{b:eq:facsigb}
\end{equation}
The short-distance coefficients $F_n$ can be calculated
using Feynman diagram methods.  Each $F_n$ can be expressed as
a perturbation series in $\alpha_s(m_Q)$. The appropriate scale for the
process is $m_{Q}$ since this
is the scale that is associated
with the production of a $Q\bar{Q}$ pair with small relative momentum.

The NRQCD matrix elements
$\langle 0|{\cal O}^{H}_n |0  \rangle$
are of the form
\begin{equation}
\langle 0|{\cal O}^{H}_n |0  \rangle =
\sum_{X}\sum_{m_J}
\; \langle 0 | {\cal K}_n | H_{m_J} + X \rangle \;
\langle H_{m_J} + X | {\cal K}'_n | 0 \rangle \; ,
\label{form:of:matrix:element}
\end{equation}
where
${\cal K}_n$ and ${\cal K}'_n$ are bilinear heavy-quark field
operators; ${\cal K}_n$
may take forms such as
$\chi^\dagger\psi$,
$\chi^\dagger\sigma^i\psi$,
$\chi^\dagger
\mbox{\boldmath$\sigma$}
\cdot \stackrel{\leftrightarrow}{{\bf D}}\psi$, {\it etc.} and
${\cal K}'_n$ takes forms
such as $\psi^\dagger\chi$, {\it etc.}
%
%
The symbol $d_n$ appearing in the factorization formula is
the combined mass dimensions of the operators ${\cal K}_n$ and ${\cal K}_n'$.
The heavy-quark field operators $\psi$ and $\chi^\dagger$ are
color-triplet columns, and are
defined to annihilate a heavy quark and anti-quark, respectively.
Thus, the ${\cal K}_n$ annihilate $Q\bar{Q}$ pairs while
the ${\cal K}'_n$ create them.
The insertion of a color matrix $T^a$ in the bilinear operators
${\cal K}_n$ and ${\cal K}'_n$,
{\it e.g.} $\chi^\dagger T^a \psi$,
means that the
bilinears project a color-octet rather than a color-singlet
$Q\bar{Q}$ state.
The index $n$ labels the various properties of the
${\cal K}_n$ and ${\cal K}'_n$.  These
properties include: 1)
the color-state of the $Q\bar{Q}$ pair projected by the operator (singlet
or octet, designated by $\underline{1}$ or $\underline{8}$);
2) the angular-momentum state of the $Q\bar{Q}$ pair projected by the
operator
(given using the spectroscopic notation ${}^{2S+1}L_J$);
and 3)  the order of the operator in the momentum expansion in the
NRQCD effective lagrangian.
The order in the momentum expansion can be increased simply by
inserting into
${\cal K}_n$ or ${\cal K}'_n$ the scalar
$(-\frac{i}{2}\stackrel{\leftrightarrow}{{\bf D}})^2/m_c^2$,
where $\stackrel{\leftrightarrow}{{\bf D}}$ is an
$SU(3)$ covariant derivative.  In
Eq.~(\ref{form:of:matrix:element}), the sum over $X$ is a symbolic
reminder that we are calculating inclusive quarkonium production.
In the discussion that follows, the sums over $X$ and
$m_J$ will always be assumed.

The factorization formula Eq.~(\ref{b:eq:facsigb})
contains an arbitarary
factorization scale $\Lambda$.
In the matrix elements
%
%
the scale $\Lambda$ can be identified with the ultraviolet cutoff of
the NRQCD effective theory.
%
Since physical results are independent of $\Lambda$,
dependence on $\Lambda$ in the short-distance
coefficients cancels against that in the NRQCD matrix elements.

In the calculation of the factorization formula for a given
quarkonium production rate, it is first necessary to determine which
are the most important terms in the series.  This operation entails
a consideration of two issues: the order in $\alpha_s(m_Q)$
of the $F_n$ and the order in $v^2$ of
the $\langle 0 | {\cal O}_n | 0 \rangle$,
{\it i.e.} their ``$v$-scaling."
In the next subsection, we will turn to the issue of
$v$-scaling of the NRQCD matrix elements.

\subsection{$v$-scaling of NRQCD matrix elements}

The most practical approach to the issue of $v$-scaling
of NRQCD matrix elements is to
define a $v$-scaling ``baseline," with all other matrix elements
being suppressed by some relative power of $v$.
This baseline corresponds to
those matrix elements in which the bilinear operators
${\cal K}_n$ and ${\cal K}'_n$ project out
the predominant $Q\bar{Q}$ component of the quarkonium, and in which this
predominant component is in an S-wave.  For charmonium, examples of such
baseline matrix elements are
\begin{eqnarray}
\langle 0 | {\cal O}_1^\eta({}^1S_0) | 0 \rangle \; & = & \;
\langle 0 | \chi^\dagger \psi |
\eta + X \rangle\;\langle \eta + X | \psi^\dagger \chi | 0 \rangle\nonumber\\
\langle 0 | {\cal O}_1^\psi({}^3S_1) | 0 \rangle  \; & = & \;
\langle 0 | \chi^\dagger
\mbox{\boldmath$\sigma$}
\psi |
\psi_{m_J}+ X \rangle
\cdot
\langle \psi_{m_J} + X |
\psi^\dagger
\mbox{\boldmath$\sigma$}
\chi | 0 \rangle \; .
\label{baseline:examples}
\end{eqnarray}
If one inserts the scalar operator
$(-\frac{i}{2}\stackrel{\leftrightarrow}{{\bf D}})^2/m_c^2$
in either of the bilinear operators, the resulting
matrix elements are suppressed by $v^2$ compared to those
in Eqs.~(\ref{baseline:examples}).

Next let us consider
those matrix elements in which the bilinear operators project out
the predominant $Q\bar{Q}$ component of the quarkonium, and in which this
predominant component is in a P-wave.  Examples are

\begin{eqnarray}
\langle 0 | {\cal O}_1^h({}^1P_1) | 0 \rangle \; & = & \;
\langle 0 |
\chi^\dagger\left( -\frac{i}{2}\stackrel{\leftrightarrow}{{\bf
D}}\right) \psi
| h_{m_J}  + X \rangle
\cdot
\langle h_{m_J} + X |
\psi^\dagger\left( -\frac{i}{2}\stackrel{\leftrightarrow}{{\bf
D}}\right)
\chi | 0 \rangle\\
\langle 0 | {\cal O}_1^{\chi_0}({}^3P_0) | 0 \rangle \;  & = & \;
\frac{1}{3}\;\langle 0 |
\chi^\dagger\left( -\frac{i}{2}\stackrel{\leftrightarrow}{{\bf D}}\cdot
\mbox{\boldmath$\sigma$}
\right) \psi
| \chi_0 + X \rangle \;
\langle \chi_0 + X |
\psi^\dagger\left( -\frac{i}{2}\stackrel{\leftrightarrow}{{\bf D}}\cdot
\mbox{\boldmath$\sigma$}
\right) \chi
| 0 \rangle\\
\langle 0 | {\cal O}_1^{\chi_1}({}^3P_1) | 0 \rangle  \; & = & \;
\frac{1}{2} \; \langle 0 |
\chi^\dagger\left( -\frac{i}{2}\stackrel{\leftrightarrow}{{\bf D}}\times
\mbox{\boldmath$\sigma$}
\right) \psi
| \chi_{1 m_J}  + X \rangle
\cdot
\langle \chi_{1 m_J} + X |
\psi^\dagger\left( -\frac{i}{2}\stackrel{\leftrightarrow}{{\bf D}}\times
\mbox{\boldmath$\sigma$}
\right) \chi
| 0 \rangle\\
\langle 0 | {\cal O}_1^{\chi_2}({}^3P_2) | 0 \rangle \; & = & \;
\langle 0 |
\chi^\dagger\left( -\frac{i}{2}\stackrel{\leftrightarrow}{D}^{[i}\sigma^{j]}
\right)  \psi
| \chi_{2 m_J} + X \rangle \;
\langle \chi_{2 m_J} + X |
\psi^\dagger\left( -\frac{i}{2}\stackrel{\leftrightarrow}{D}^{[i}\sigma^{j]}
\right) \chi
| 0 \rangle \; .
\end{eqnarray}
To a rough approximation, the above matrix elements, when divided by
$m_c^2$, are
suppressed with respect to the baseline by $v^2$, and of course, the insertion
of
$(-\frac{i}{2}\stackrel{\leftrightarrow}{{\bf D}})^2/m_c^2$
would result in further suppression by $v^2$.

The preceding discussion has merely involved counting powers of momentum.
There is, however, an important additional issue in the $v$-scaling of
NRQCD matrix elements.
This second issue
hinges on the Fock state expansion of the quarkonium meson.
The Fock state expansion of, for example, the $\psi$ particle, in
Coulomb gauge,
can be thought of schematically as
\begin{eqnarray}
|\, \psi \, \rangle =
A_{c\bar{c}}    |c\bar{c}(\underline{1},{}^3S_1) \,\rangle
+ A_{c\bar{c}g} |c\bar{c}(\underline{8},{}^3P_J) \,  g \rangle
+ A_{c\bar{c}gg}|c\bar{c}(\underline{8},{}^3S_1) \,  g g \rangle
+ B_{c\bar{c}gg}|c\bar{c}(\underline{1},{}^3S_1) \,  g g \rangle\nonumber\\
+ C_{c\bar{c}gg}|c\bar{c}(\underline{8},{}^3D_J) \,  g g \rangle
+ D_{c\bar{c}gg}|c\bar{c}(\underline{1},{}^3D_J) \,  g g \rangle
+ B_{c\bar{c}g} |c\bar{c}(\underline{8},{}^1S_0) \,  g \rangle + \cdots,
\label{fockstate}
\end{eqnarray}
where $g$ represents a dynamical gluon, {\it i.e.} one whose effects cannot be
incorporated into an instantaneous potential and whose typical momentum is
$m_c v^2$.    The angular-momentum
quantum numbers of the $c\bar{c}$ pairs in the various Fock states
are indicated in spectroscopic notation, and their color
configurations are labeled by $\underline{1}$ for singlet or
$\underline{8}$ for octet.
We will now discuss the $v$-scaling of the various coefficients $A$, $B$,
{\it etc.}

Since the state $| c\bar{c}(\underline{1},{}^3S_1)\rangle$
is the predominant Fock state in $|\psi\rangle$, we expect that the
coefficient $A_{c\bar{c}}$ is just a little less than unity, {\it i.e.}
$A_{c\bar{c}}\sim v^0$.  As to the Fock state
$|c\bar{c}(\underline{8},{}^3P_J)g\rangle$,
this configuration arises when the predominant state radiates a soft dynamical
gluon; such a process is mediated principally by the electric dipole operator,
for which the selection rule is $L' = L\pm 1$, $S' = S$, and which involves
a single power of heavy quark three-momentum; thus, the coefficient
$A_{c\bar{c}g}$ is of order $v^1$.  The electric dipole emission of yet
another gluon involves a change from the $P$-wave state
$A_{c\bar{c}g}|c\bar{c}(\underline{8},{}^3P_J)g\rangle$
to the $S$- and $D$-wave states
$|c\bar{c}(\underline{8},{}^3S_1) \,  g g \rangle$,
$|c\bar{c}(\underline{1},{}^3S_1) \,  g g \rangle$,
$|c\bar{c}(\underline{8},{}^3D_J) \,  g g \rangle$ and
$|c\bar{c}(\underline{1},{}^3D_J) \,  g g \rangle$;
their coefficients ---
$A_{c\bar{c}gg}$,
$B_{c\bar{c}gg}$,
$C_{c\bar{c}gg}$ and
$D_{c\bar{c}gg}$ --- are of order $v^2$.  Lastly, we consider the coefficient
of the state $| c\bar{c}(\underline{8},{}^1S_0)g\rangle$.  Fluctuations
into this spin-singlet state from the predominant spin-triplet state
are associated with the emission of a soft gluon via a spin-flipping
magnetic dipole transition; such transitions involve
the gluon three-momentum
$(\sim m_c v^2)$
rather than the heavy quark three-momentum
$(\sim m_c v^1)$,
and therefore the coefficient $B_{c\bar{c}g}$ is of order $v^2$.

We now bring together the
power-counting rules and the Fock state ideas presented above.  Let us
consider
as an example the matrix element
\begin{equation}
\langle 0 | {\cal O}_8^\psi({}^3S_1) | 0 \rangle =
\langle 0 | \chi^\dagger \sigma^i T^a\psi |
\psi_{m_J}+ X \rangle
\langle \psi_{m_J} + X |
\psi^\dagger \sigma^i T^a\chi | 0 \rangle .
\end{equation}
The bilinear operators project out the
$|c\bar{c}(\underline{8},{}^3S_1)gg\rangle$ Fock state, whose coefficient
in the expansion is of order $v^2$.  Thus,
$\langle 0 | O_8^\psi({}^3S_1) | 0 \rangle$
is suppressed by roughly $v^4$ compared to the $S$-wave baseline.

As another example, consider the $v$-scaling of the matrix element
$\langle 0 | {\cal O}_8^\psi({}^3P_J) | 0 \rangle$.  The bilinear operators
contained therein project out the state
$|c\bar{c}(\underline{8},{}^3P_J)g\rangle$, whose coefficient
in the expansion would be of order $v^1$.  This Fock state suppression,
combined with the derivatives inherent in the $P$-wave
operators, gives a $v$-scaling of $v^4$ (with respect to the baseline) for
$\langle 0 | {\cal O}_8^\psi({}^3P_J) | 0 \rangle$.

The $v$-scaling of those NRQCD matrix elements pertinent to the calculation
of hadronic $\psi$ production
are given in Table I.

\subsection{Order in $\alpha_s$ of short-distance coefficients}

The leading-order Feynman diagrams for the production of
a $c\bar{c}$ pair in  color-octet
states are presented in Figures 2 and 3.
The diagram in Figure 2 ($q\bar{q}\to g\to c\bar{c}$)
produces a $c\bar{c}(\underline{8},{}^3S_1)$ state.
The diagrams in Figure 3a (the gluon fusion process $gg\to c\bar{c}$) produce
$c\bar{c}(\underline{8},{}^1S_0)$,
$c\bar{c}(\underline{8},{}^3S_1)$
and $c\bar{c}(\underline{8},{}^3P_J)$ states.
The diagram in Figure 3b ($gg\to g \to c\bar{c}$) produces a
$c\bar{c}(\underline{8},{}^3S_1)$ state.

It is interesting to note that,
at leading order in the ${\bf q}^2/m_c^2$ expansion, the
amplitude for the production of $c\bar{c}(\underline{8},{}^3S_1)$
from Figures 3a cancels against the amplitude for the
same quantum number production in Figure 3b, so that the
total production of $c\bar{c}(\underline{8},{}^3S_1)$
from gluon-gluon collisions vanishes.

To determine the rough size of a term in the factorization formula,
one attributes to it a power of $\alpha_s(m_c) \approx 0.3$ for each
vertex in the Feynman diagram, and the appropriate power
of $v^2 \approx 0.3$ according to its $v$-scaling.
We summarize the $\alpha_s(m_c)$ powers associated
with the leading order subprocesses
in Table II.

Once the most important terms in the factorization formula have
been identified, one procedes to compute the short-distance coefficients
$F_n$.  This is done by matching perturbative full QCD and perturbative
NRQCD calculations for the production rate of
$c\bar{c}$ pairs with specific color and angular-momentum quantum numbers.
These rates are
expressed as an integral over
relative momentum ${\bf q}$, with the integrand given
as a Taylor expansion in ${\bf q}^2/m_c^2$

In the next sections, we calculate the production rates for the various
important octet subprocesses contributing to hadronic $\psi$ production,
and outline in detail the matching procedure.

\section{Production Rate for \lowercase{$ q\bar{q} \to \psi $} Subprocess}

We now calculate the leading-order
contribution to the cross section
for inclusive $\psi$ production from the color-octet
subprocess $q \bar{q} \to \psi$. This is done in three
steps. First we carry out a perturbative QCD calculation of
the production rate
for the process $q(k) \bar{q}(\bar{k}) \to c(p) \bar{c}(\bar{p})$ with
on-shell heavy quarks.
The second step consists
of calculating the
production rate for the same process in NRQCD.
The final step will be to determine the short-distance
coefficient $F_n$ by matching the QCD and NRQCD calculations.

\subsection{Production rate in full QCD}

In noncovariant conventions, the transition amplitude for the process
$ q(k) \bar{q}(\bar{k}) \to c(p) \bar{c}(\bar{p})$, illustrated in
Figure~2, is given by
\begin{equation}
{\cal T} \; = \;
{g^2_s \over 4 \sqrt{k_0\bar{k}_0p_0\bar{p}_0} } \; \bar{v}(\bar{k})
\gamma^{\mu} T^a_{\bar{l}l} u(k) \; {1 \over (p+\bar{p})^2} \;
\bar{u}_s(p) \gamma_{\mu} T^a_{m\bar{m}} v_{\bar{s}}(\bar{p}) \; ,
\label{qqbar}
\end{equation}
where $l$, $\bar{l}$, $m$ and $\bar{m}$ are color indices and
$s$ and $\bar{s}$ are heavy-quark spin indices.  The matrices
$T^a$ are normalized throughout so that $Tr[ T^a T^b] = \delta^{ab}/2$.
It is convenient
to re-express the $c$ and $\bar{c}$ momentum in terms of ${\bf q}$
(the relative three-momentum of the quark and antiquark in
the $c \bar{c}$ rest frame), and $P$ (the total four-momentum
of the quark and anti-quark in the lab frame):
\begin{eqnarray}
p^{\mu} & = & {1 \over 2} P^{\mu} \; + \; \Lambda^{\mu i}(P)q^i,
\nonumber \\
\bar{p}^{\mu} & = & {1 \over 2} P^{\mu} \; - \; \Lambda^{\mu i}(P)q^i,
\label{mom}
\end{eqnarray}
where the Lorentz boost
matrix $\Lambda^{\mu \nu}(P)$ is given by
\begin{eqnarray}
\Lambda^0_j & = & \frac{1}{2E_q} P^j\nonumber\\
\Lambda^i_j & = & \delta^{ij} -
\frac{P^iP^j}{{\bf P}^2}
+ \frac{P^0}{2E_q}\,
\frac{P^iP^j}{{\bf P}^2}\; ,
\end{eqnarray}
with
$E^2_q = m^2_c+{\bf q}^2$. The boost matrix has the following useful
properties:
\begin{eqnarray}
\Lambda^{\mu}_{i} \Lambda^{\nu}_{j} \delta_{ij} & = &
-g^{\mu \nu}+{ P^{\mu}P^{\nu} \over P^{2}}
\nonumber \\
\Lambda^{\mu}_{i} \Lambda^{\nu}_{j} g_{\mu \nu} & = &
-\delta_{ij} \; .
\label{boostid}
\end{eqnarray}
The heavy quark Dirac
bilinear $\bar{u}(p)\gamma^{\mu} v(\bar{p})$ appearing in Eq.~(\ref{qqbar})
is now re-expressed in
terms of ${\bf q}$,
the Pauli two spinor $\xi$ and antispinor $\eta$,
and the Pauli matrices $\sigma^i$~\cite{bc}:
\begin{equation}
\bar{u}_s(p) \gamma^{\mu} v_{\bar{s}}(\bar{p}) =
\Lambda^{\mu}_j \left( 2E_q \; \xi^{\dagger}_s \sigma^j \eta_{\bar{s}} -
{2 \over E_q +m_c} \; q^j
\xi^{\dagger}_s {\bf q} \cdot \mbox{\boldmath$\sigma$} \eta_{\bar{s}}
\right)\; ,
\label{red1}
\end{equation}
where $\bar{u}_s(p)u_s(p) = 2m_c$
and $\xi^{\dagger}_s \xi_t = \delta_{st}$.
The procedure of re-expressing the manifestly Lorentz-invariant
Dirac bilinear current in terms of two-spinors can be called
the ``reduction of Dirac bilinears.''
Applying the reduction to Eq.~(\ref{qqbar}) we obtain
\begin{eqnarray}
{\cal T} & = & {g^2_s \over 2
\sqrt{k_0\bar{k}_0(P^2_0-4({\bf \Lambda_0} \cdot {\bf q})^2)}}
\;\bar{v}(\bar{k})
\gamma^{\mu} T^a_{\bar{l}l} u(k) \; {1 \over 4 E^2_q}
\nonumber \\
& & \; \; \; \times
\Lambda_{j \mu}(P) \left( 2E_q \xi^{\dagger}_{s} \sigma^j
T^a_{m\bar{m}}\eta_{\bar{s}}  - {2 \over E_q + m_c} q^j
\xi^{\dagger}_s
{\bf q} \cdot
\mbox{\boldmath$\sigma$} T^a_{m\bar{m}}\eta_{\bar{s}} \right) \; .
\label{rodb}
\end{eqnarray}
(In the above context,
the symbol $T^a_{m\bar{m}}$ represents a number, not a matrix;
in our notation, the two-spinors $\xi$ and $\eta$ do not carry a color index
and are not acted upon by the $T^a$.
Our strategy of placing the factor $T^a_{m\bar{m}}$ between
the two-spinors is designed to make the matching procedure
more transparent.)
The bilinear combination $\xi^{\dagger}_s \sigma_j \eta_{\bar{s}}$
corresponds to
a ${}^3S_1$ spin configuration of the $c \bar{c}$ pair.  Working
to leading order in ${\bf q}$, we keep only
the ${}^3S_1$
term:
\begin{equation}
{\cal T} \; = \;
{g^2_s \over 2 \sqrt{k_0\bar{k}_0(P^2_0-4({\bf \Lambda_0} \cdot {\bf q})^2)}}
\; \bar{v}(\bar{k})
\gamma^{\mu} T^a_{\bar{l}l} u(k) \; {1 \over 2 E_q} \;
\Lambda_{j \mu}(P) \; \xi^{\dagger}_s \sigma^j T^a_{m\bar{m}}
\eta_{\bar{s}} \; .
\label{redqqbar}
\end{equation}
Anticipating that
the probability of hadronization is negligible for
$|{\bf q}| > \Lambda_{QCD}$,
we need only derive the algebraic form of ${\cal T}$ for small ${\bf q}$,
and therefore need only keep the first few
terms in the Taylor expansion in ${\bf q}^2/m_c^2$:
\begin{equation}
{\cal T} \; \approx \;
{g^2_s \over 2 P_0 \sqrt{k_0 \bar{k}_0}}
\;\bar{v}(\bar{k}) \gamma^{\mu} T^a_{\bar{l}l} u(k) \; {1 \over 2 m_c} \;
\Lambda_{j \mu} \;  \xi^{\dagger}_s \sigma^j T^a_{m\bar{m}} \eta_{\bar{s}}
\left( 1 + a \frac{{\bf q}^2}{m_c^2} + \cdots \right) \; ,
\label{nrqqbar}
\end{equation}
where the constant $a$ --- which we do not bother
to calculate explicitly  --- is
the coefficient of the second term in the Taylor series
expansion in ${\bf q}^2/m_c^2$.  It must be kept in mind
that the factor appearing in brackets is not, at this point
in the calculation, a rapidly converging series.  Indeed,
the upper bound on ${\bf q}$ is associated with the
total momentum available in the experiment, and in general this can
be much greater than $m_c$.  However, for the purposes
of matching in the factorization formalism,
we treat ${\cal T}$ as an expansion in ${\bf q}^2/m_c^2$,
and consider only the first few terms, the probability of hadronization
being non-negligible only for small ${\bf q}$.

Multiplying~(\ref{nrqqbar}) by its complex conjugate,
summing over final colors and spins, and
averaging over initial colors and spins, we obtain
\begin{equation}
|\bar{{\cal T}}|^2 \; \approx \;
{g^2_s \over 54(2m_c)^4 } \;
\xi^{\dagger}_s
\mbox{\boldmath$\sigma$}
T^a_{m\bar{m}} \eta_{\bar{s}} \cdot
\eta^{\dagger}_{\bar{s}}
\mbox{\boldmath$\sigma$}
T^a_{\bar{m}m} \xi_{s}\,
\left( 1 + 2 a \frac{{\bf q}^2}{m_c^2} + \cdots \right) \; ,
\label{probqqbar}
\end{equation}
where a sum over spins $s$ and $\bar{s}$
and color $m$ and $\bar{m}$  is assumed.
Here we have made use of the fact that
\begin{equation}
\sum_{s\bar{s}}
\;\xi^{\dagger}_s
\sigma^i
\eta_{\bar{s}} \;\;
\eta^{\dagger}_{\bar{s}}
\sigma^j
\xi_s
= { \delta^{ij} \over 3} \; \sum_{s\bar{s}} \;
\xi^{\dagger}_s
\mbox{\boldmath$\sigma$}
\eta_{\bar{s}} \cdot
\eta^{\dagger}_{\bar{s}}
\mbox{\boldmath$\sigma$}
\xi_s \; .
\label{psid1}
\end{equation}

The expression for the cross section in terms of the transition
amplitude squared is
\begin{equation}
\sigma  = {1 \over F} \;  \int {d^3 {\bf p} \over (2\pi)^3}
\int {d^3 {\bf\bar{p}} \over (2\pi)^3} \; \;
(2 \pi)^4 \delta^4(P-k-\bar{k})
\; |\bar{{\cal T}}|^2 \; ,
\label{dcs}
\end{equation}
where $F= 2$ is the noncovariant flux. Changing variables
from ${\bf p}$ and ${\bf \bar{p}}$ to
${\bf P}$ and ${\bf q}$,
the above expression becomes
%
%
\begin{eqnarray}
\sigma  &=& {1 \over F} \;  \int {d^3 {\bf P} \over (2\pi)^3}
\int {d^3 {\bf q} \over (2\pi)^3} \; {P^0 \over 2 E_q} \;
(2 \pi)^4 \delta^4(P-k-\bar{k})
\; |\bar{{\cal T}}|^2  \;
\nonumber\\
%
%
&=& \pi  \int {d^3 {\bf q} \over (2\pi)^3} \; \;
\delta(E_f - E_i) \; |\bar{{\cal T}}|^2 \;
\left( 1 +  a' \frac{{\bf q}^2}{m_c^2} + \cdots \right).
\label{crsc}
\end{eqnarray}

Inserting Eq.~(\ref{probqqbar}) into Eq.~(\ref{crsc}) we obtain the
rate for on-shell $c \bar{c}$ production:
\begin{equation}
\sigma(q \bar{q} \to c \bar{c}(\underline{8},{}^3S_1))
= {\alpha^2_s \pi^3 \over 54 m^4_c} \;
\int {d^3 {\bf q} \over (2\pi)^3} \;
\delta( E_f - E_i) \;
\xi^{\dagger}_{s}
\mbox{\boldmath$\sigma$}
T^a_{m\bar{m}} \eta_{\bar{s}}\;\;
\eta^{\dagger}_{\bar{s}}
\mbox{\boldmath$\sigma$}
T^a_{\bar{m}m} \xi_{s}\;
\left( 1 +  a'' \frac{{\bf q}^2}{m_c^2} + \cdots \right).
\label{csqqbar}
\end{equation}
%


\subsection{Production rate in NRQCD}

The next step is to calculate the production rate
for the process $q \bar{q} \to c \bar{c}(\underline{8},{}^3S_1)$ in
perturbative NRQCD.
The first two terms are
\begin{equation}
\sigma(q \bar{q} \to c \bar{c}(\underline{8},{}^3S_1))=
\int\frac{d^3 {\bf q}}{(2\pi)^3}
\left(
\;\;
\frac{F_8({}^3S_1)}{m_c^2} \;\;
\langle 0 | {\cal O}^{c \bar{c}(q)}_8({}^3S_1) | 0 \rangle
\; +
\; \frac{G_8({}^3S_1)}{m_c^4}\;\;
\langle 0 | {\cal P}^{c \bar{c}(q)}_8({}^3S_1) | 0 \rangle
+ \cdots \right) \; ,
\label{nrqcd:form:1}
\end{equation}
where $F_8({}^3S_1)$ and $G_8({}^3S_1)$ are specific cases of
the short-distance coefficients $F_n$, and where
\begin{eqnarray}
\langle 0 | {\cal O}^{c \bar{c}(q)}_8({}^3S_1) | 0 \rangle & = &
\sum_{s\bar{s}m\bar{m}}
\langle 0 | \chi^\dagger \sigma^i T^a \psi
| c(s,m,{\bf q}) \bar{c}(\bar{s},\bar{m}, -{\bf q}) \rangle\nonumber\\
&& \qquad\langle
c(s,m,{\bf q}) \bar{c}(\bar{s},\bar{m},-{\bf q})|
\psi^\dagger \sigma^i T^a \chi | 0 \rangle
\label{line:1}
\\
\langle 0 | {\cal P}^{c \bar{c}(q)}_8({}^3S_1) | 0 \rangle & = &
\frac{1}{2} \sum_{s\bar{s}m\bar{m}}
\langle 0
| \chi^\dagger \sigma^i T^a
\left(-\frac{i}{2}\stackrel{\leftrightarrow}{{\bf D}}\right)^2
 \psi
| c(s,m,{\bf q}) \bar{c}(\bar{s},\bar{m}, -{\bf q}) \rangle\nonumber\\
&& \qquad\langle
c(s,m,{\bf q}) \bar{c}(\bar{s},\bar{m},-{\bf q})|
\psi^\dagger \sigma^i T^a \chi
| 0 \rangle\nonumber\\
& + &
\frac{1}{2} \sum_{s\bar{s}m\bar{m}}
\langle 0
| \chi^\dagger \sigma^i T^a
 \psi
| c(s,m,{\bf q}) \bar{c}(\bar{s},\bar{m}, -{\bf q}) \rangle\nonumber\\
&&\qquad\langle
c(s,m,{\bf q}) \bar{c}(\bar{s},\bar{m},-{\bf q})|
\psi^\dagger \sigma^i T^a
\left(-\frac{i}{2}\stackrel{\leftrightarrow}{{\bf D}}\right)^2
\chi
| 0 \rangle \; .
\label{line:2}
\end{eqnarray}

Our task here is to derive explicit expressions for the $c\bar{c}(q)$
matrix elements in Eqs.~(\ref{line:1}) and (\ref{line:2}).
For practical reasons, it is worthwhile to quickly note
one possible set of
conventions for these calculations.
Let us define the single-particle annihilation and creation operators
to obey the anticommutation relation
$[a(s,i,{\bf q}),a^\dagger(t,j,{\bf p})] =
(2\pi)^3 \delta({\bf q} - {\bf p}) \delta^{st}\delta^{ij}$.
The annihilation operator acts according to
$a(s,i,{\bf q})| c(t,j,{\bf p})\rangle =
(2\pi)^3
\delta({\bf q} - {\bf p}) \delta^{st}\delta^{ij}
|0\rangle$,
and the creation operator according to
$a^\dagger(t,j,{\bf q}) |0\rangle = | c(t,j,{\bf q})\rangle $.
Single particle states are normalized so that
$\langle c(s,i,{\bf q})| c(t,j,{\bf p})\rangle
= (2\pi)^3 \delta({\bf q} - {\bf p})\delta^{st}\delta^{ij}$.
Then,
conceiving of the field operators $\psi$ and $\chi$
as color-triplet vectors, we write their
Fourier decompositions as
\begin{eqnarray}
\psi_m(x) & = & \sum_s \int {d^3 p \over (2 \pi)^3}
a(s,m,{\bf p}) \xi_s e^{-ip\cdot x }
\nonumber \\
\chi^{\dagger}_{\bar{m}}(x) & = & \sum_{\bar{s}} \int {d^3 p \over (2 \pi)^3}
b(\bar{s},\bar{m},{\bf p}) \eta^{\dagger}_{\bar{s}} e^{-ip\cdot x } \; ,
\label{pwdec}
\end{eqnarray}
where $\xi_s$ and $\eta^{\dagger}_{\bar{s}}$ are the two-spinors.

Using the conventions outlined in the preceding paragraph,
we derive explicit expressions
for the right-hand-side of Eq.~(\ref{nrqcd:form:1}), obtaining
\begin{equation}
\sigma(q \bar{q} \to c \bar{c}(\underline{8},{}^3S_1) )=
\int\frac{d^3{\bf q}}{(2\pi)^3}
\left(
\;\;\frac{F_8({}^3S_1)}{m_c^2}\; + \;
\frac{G_8({}^3S_1)}{m_c^2}\; \frac{{\bf q}^2}{m_c^2} + \cdots \right)
\; \xi^{\dagger}_{s}
\mbox{\boldmath$\sigma$}
T^a_{m\bar{m}} \eta_{\bar{s}} \;\;
\eta^{\dagger}_{\bar{s}}
\mbox{\boldmath$\sigma$}
T^a_{\bar{m}m} \xi_{s} \; .
\label{ffnrqcdsig}
\end{equation}

\subsection{Matching}

Finally we determine the short-distance coefficient by matching
the perturbative QCD result Eq.~(\ref{csqqbar}) to the perturbative NRQCD
result Eq.~(\ref{ffnrqcdsig})
\begin{eqnarray}
F_8({}^3S_1)  & = &
{2 \alpha^2_s \pi^3 \over 27 m_c} \; \delta (s- 4m^2_c)\\
G_8({}^3S_1)  & = &
{2 a'' \alpha^2_s \pi^3 \over 27 m_c} \; \delta (s - 4m^2_c) \; .
\label{sdc}
\end{eqnarray}
The matching procedure for on-shell scattering fixes the short-distance
coefficient in the non-relativistic effective theory. However, these
same coefficients apply to the calculation of the production of
bound quarkonium states.  This is because the non-perturbative
effects that will bind the $c$ and $\bar{c}$ into a charmonium meson
take place
over much longer distances than the separation $1/m_c$ associated with
the formation of the $c\bar{c}$ pair. Thus, although the
short-distance coefficient has been determined using a perturbative
calculation of the production of free quarks and antiquarks, the same
short-distance coefficient applies to the formation of a $c\bar{c}$
pair in a charmonium state. The nonperturbative effects involved in
the formation of the boundstate are described by the matrix elements.
Therefore, the
inclusive cross section for $q \bar{q} \to \psi$ via the color-octet
mechanism is, to leading order in $v^2$,
\begin{eqnarray}
\sigma(q \bar{q} \to \psi) & = &
\frac{F_8({}^3S_1)}{m_c^2}
\langle 0 | {\cal O}_8^\psi({}^3S_1) | 0 \rangle\nonumber\\
&=&{ 2 \alpha^2_s \pi^3 \over 27 m^3_c} \; \delta (s - 4 m^2_c)
 \langle 0|{\cal O}^{\psi}_8 ({}^3S_1) | 0 \rangle \; .
\label{psics}
\end{eqnarray}

This procedure, which involves the determination of
the $F_n$ for the production of on-shell heavy quarks,
correctly takes into account the effects of binding energy,
as can be verified by explicit calculations in
NRQED~\cite{pl}.

\section{Cross section for $\lowercase{gg} \to \psi $ subprocess}

Next we turn our attention to the
process $g(g_1) g(g_2) \to c(p) \bar c(\bar p)$ which, at leading
order in $\alpha_s$,
proceeds through the Feynman diagrams in Figure~3.
To determine the short-distance coefficients we follow
a sequence of steps analogous to that outlined
in the previous section. At this point we will abandon or habit of
writing heavy quark spin and color indices explicitly.

The transition amplitude calculated
from the diagrams in Figure~3a is
\begin{eqnarray}
\lefteqn{ {\cal T}
(gg \to c\bar{c})
={- g^2_s\over 4\sqrt{g_{10}g_{20}p_0\bar{p}_0 } }
\; \epsilon_\mu(g_1) \epsilon_\nu(g_2)   }
\nonumber \\
& & \; \; \; \bar u(p)
\left[ T^a T^b
{\gamma^\mu ( \not \! p - \not \! g_1 + m_c) \gamma^\nu
	\over 2 p \cdot g_1} \right.
+ T^b T^a {\gamma^\nu (- \not \! \bar{p} + \not \! g_1 + m_c)
\gamma^\mu \over 2 \bar{p} \cdot g_1} \left. \right] v(\bar p) \; .
\label{mtrx_elem}
\end{eqnarray}
The product of color matrices in Eq.~(\ref{mtrx_elem}) can be rewritten as
\begin{equation}
T^a T^b = {1 \over 6} \delta^{ab} + {1 \over 2}(d^{abc}+if^{abc})T^c,
\label{color_mtrx}
\end{equation}
where $d^{abc}$ is a symmetric tensor and $f^{abc}$ is an
antisymmetric tensor. Since we wish to determine the short-distance
coefficient for color-octet production, we will discard the $\delta^{ab}/6$
term of Eq.~(\ref{color_mtrx}). Then, using the Dirac equation,
Eq.~(\ref{mtrx_elem}) can
be transformed into
\begin{eqnarray}
{\cal T}(gg\to c\bar{c})  & = &
{g^2_s \over 16 E_q^2 \sqrt{g_{10}g_{20} p_0 \bar{p}_0} }
\left[ 1 - {(g_1 \cdot \Lambda q)^2 \over E_q^4} \right]^{-1}
\; \epsilon_\mu(g_1) \epsilon_\nu(g_2)
\nonumber \\
& &  \; \; \; \; \bar u(p) \Bigg\{ d^{abc} T^c
\left[ {\cal A}^{\mu \nu} -
{g_1 \cdot \Lambda q \over E_q^2}{\cal B}^{\mu \nu} \right]
\; + \; if^{abc}T^c \left[ {\cal B}^{\mu \nu} -
{g_1 \cdot \Lambda q \over E_q^2}{\cal A}^{\mu \nu} \right] \Bigg\}
v(\bar{p}) \; ,
\nonumber \\
\label{red_me}
\end{eqnarray}
where
\begin{equation}
{\cal A}^{\mu \nu} = i \epsilon^{\mu \nu \alpha \beta} (g_1 - g_2)_\alpha
\gamma_\beta \gamma_5
\; + \; 2 \left[ (\Lambda q)^\mu \gamma^\nu +
(\Lambda q)^\nu \gamma^\mu \right] \; ,
\label{A}
\end{equation}
and
\begin{equation}
{\cal B}^{\mu \nu} =  2 g^\mu_2 \gamma^\nu - 2 g^\nu_1 \gamma^\mu
	+ g^{\mu \nu} (\not \! g_1 - \not \! g_2) \;  .
\label{B}
\end{equation}
We reduce
the Dirac bilinears using Eq.~(\ref{red1}) and
\begin{equation}
\bar{u}(p) \gamma^\mu \gamma_5 v(\bar{p}) =
{m_c \over E_q}P^\mu \; \xi^{\dagger} \eta
- 2i \Lambda^\mu_j \xi^\dagger
({\bf q} \times \mbox{\boldmath$\sigma$})^j \eta \; .
\label{red2}
\end{equation}
Keeping only terms of leading order in ${\bf q}$, the
resulting amplitude may be expressed in the form
\begin{equation}
{\cal T}(gg\to c\bar{c})
\approx {\cal T}_{CS}(gg\to c\bar{c}) +
{\cal T}_{CAS} (gg\to c\bar{c}) \; ,
\label{amp12}
\end{equation}
where ${\cal T}_{CS}$ is the color symmetric piece
\begin{eqnarray}
{\cal T}_{CS}(gg\to c\bar{c})  & = & d^{abc}
{g^2_s \over (2m_c)^4}
\epsilon_\mu(g_1) \epsilon_\nu(g_2)
\Bigg\{ i \epsilon^{\mu \nu \alpha \beta}
(g_1 - g_2)_{\alpha} P_{\beta} \;
\xi^{\dagger} T^c \eta
\nonumber \\
& + & \left[ 2 \epsilon^{\mu \nu \alpha \beta}
(g_1 - g_2)_{\alpha} \epsilon^{ijk}\Lambda^k_{\beta}
+ 4m_c (\Lambda^{\mu}_i \Lambda^{\nu}_j
+ \Lambda^{\nu}_i \Lambda^{\mu}_j ) \right.
\nonumber \\
 -  {2 \over m_c} (g_1 \cdot \Lambda)_i ( 2g^{\mu}_2 \Lambda^{\nu}_j
& - & 2g^{\nu}_1 \Lambda^{\mu}_j + g^{\mu \nu} (g_1 - g_2)\cdot \Lambda_j)
\left.  \right]  q^i \xi^{\dagger} \sigma^j T^c \eta
\Bigg\} \;  + \cdots \;  ,
\label{t1red}
\end{eqnarray}
and where ${\cal T}_{CAS}$ is the color antisymmetric part
\begin{eqnarray}
{\cal T}_{CAS}(gg\to c\bar{c}) & = & i f^{abc} {g^2_s \over (2m_c)^3}
\epsilon_\mu(g_1) \epsilon_\nu(g_2) \nonumber\\
&& \qquad
\left[2 g^{\mu}_2 \Lambda^{\nu}_j - 2 g^{\nu}_1 \Lambda^{\mu}_j +
g^{\mu \nu} (g_1 - g_2 ) \cdot \Lambda_j \right]
\xi^{\dagger} \sigma^j T^c \eta \;
 + \cdots  \; .
\label{t2red}
\end{eqnarray}
The dots $ \cdots $ are included as a reminder that we
are writing only the first term in the infinite and non-convergent
series in ${\bf q}^2/m_c^2$.  This is justified since
the factorization formalism
requires only that we know the algebraic form of
${\cal T}$ for small ${\bf q}$.

We note that at leading order in the non-relativistic expansion,
the amplitude for producing $c\bar{c}(\underline{8},{}^3S_1))$ in
Figure 3b, Eq.~(\ref{t2red}) cancels against
the contribution from the
gluon-fusion graphs in Figure 3a,
leaving only the pieces in Eq.~(\ref{t1red}).

We now turn to the task of isolating the terms in Eq.~(\ref{t1red})
that contribute
to the production of $c\bar{c}$ pairs in the states
$\underline{8},{}^1S_0$, $\underline{8},{}^3P_0$
and $\underline{8},{}^3P_2$.
Beginning with $\underline{8},{}^1S_0$, we note that
terms associated with the angular-momentum
quantum numbers ${}^1 S_0$ will have Pauli-spinor bilinears of the
form $\xi^\dagger \eta$, ${\bf q}^2 \xi^\dagger \eta$, {\it etc.}  It is
an easy matter to pick out the
${}^1S_0$ pieces in
Eq.~(\ref{t1red}):
\begin{equation}
{\cal T}(gg \to c \bar{c}(\underline{8},{}^1 S_0)) = i d^{abc}
{g^2_s \over (2m_c)^4} \; \epsilon_\mu(g_1) \epsilon_\nu(g_2)
\; \epsilon^{\mu \nu \alpha \beta} (g_1 - g_2)_{\alpha}P_{\beta} \;
\; \xi^{\dagger} T^c \eta \; \;(1 + \cdots )\; .
\label{1s0}
\end{equation}
Squaring Eq.~(\ref{1s0}), and performing the sum-average over
spins and
colors, we obtain
\begin{equation}
|\bar{{\cal T}} (gg \to c\bar{c}(\underline{8},{}^1 S_0))|^2 =
{5 g^4_s \over 384 (2m_c)^4} \; \; \xi^{\dagger} T^c \eta
\;\; \eta^{\dagger} T^c \xi \;\;( 1 + \cdots ) \;  .
\label{sq1s0}
\end{equation}
Inserting this into the expression for the cross section given in
Eq.~(\ref{crsc}) we obtain
\begin{equation}
\sigma(gg \to c\bar{c}(\underline{8},{}^1 S_0)) \approx
{5 \pi^3 \alpha^2_s \over 3 \cdot 128 \, m_c^4} \;
\int {d^3 {\bf q} \over (2 \pi)^3} \;
\delta(E_f - E_i)
\;\;\xi^{\dagger} T^c \eta\;
\; \eta^{\dagger} T^c \xi \;\;( 1 + \cdots ) \; ,
\label{3s1cs}
\end{equation}
which is the result in full QCD for
the production rate of an on-shell
$c\bar{c}(\underline{8},{}^1S_0)$.

The next step consists of calculating the same production
rate in perturbative NRQCD.
This is given by
\begin{eqnarray}
\sigma(gg \to c\bar{c}(\underline{8},{}^1 S_0)) &=&
\int \frac{d^3{\bf q}}{(2\pi)^3}
\;
\frac{F_8({}^1S_0)}{m_c^2}  \;
\langle 0 | {\cal O}^{c\bar{c}(q)}_8({}^1S_0)
| 0 \rangle + \cdots \nonumber\\
& = &
\int \frac{d^3{\bf q}}{(2\pi)^3}
\;
\frac{F_8({}^1S_0)}{m_c^2}  \;
\sum_{s\bar{s}m\bar{m}}
\langle 0
| \chi^\dagger T^a
 \psi
| c(s,m,{\bf q}) \bar{c}(\bar{s},\bar{m}, -{\bf q}) \rangle\nonumber\\
&&\qquad\langle
c(s,m,{\bf q}) \bar{c}(\bar{s},\bar{m},-{\bf q})|
\psi^\dagger T^a
\chi
| 0 \rangle + \cdots \nonumber\\
& = &
\int {d^3 {\bf q} \over (2 \pi)^3}
\; \frac{F_8({}^1 S_0)}{m_c^2}
\; \;\xi^{\dagger} T^c \eta \;
\; \eta^{\dagger} T^c \xi \; + \cdots \; .
\label{nrqcd3s1cs}
\end{eqnarray}
Matching the result from full QCD in Eq.~(\ref{3s1cs}) to that from
NRQCD in Eq.~(\ref{nrqcd3s1cs}), we obtain
\begin{equation}
F_8({}^1S_0) =
\frac{5\pi^3\alpha_s^2}{96 m_c}
\delta(s - 4 m_c^2) \; ,
\end{equation}
from which we conclude that
\begin{eqnarray}
\sigma(gg \to \psi + X) & = &
\frac{F_8({}^1S_0)}{m_c^2}
\langle 0 | {\cal O}^{\psi}_8 ({}^1S_0) | 0 \rangle\nonumber\\
& = &
{5 \pi^3 \alpha^2_s \over  96 m_c^3} \; \delta (s - 4m^2_c) \;
\langle 0 | {\cal O}^{\psi}_8 ({}^1S_0) | 0 \rangle \; .
\label{psi1s0}
\end{eqnarray}

Next we isolate the individual $P$-wave contributions. This can be
accomplished by first noting that any direct product of cartesian vectors may
be written as
\begin{equation}
a^ib^j = {1 \over 3}\delta^{ij} \; {\bf a} \cdot {\bf b}
+ {1 \over 2} \epsilon^{ijk}
({\bf a} \times {\bf b})^k + a^{[i}b^{j]} \; ,
\label{decomp}
\end{equation}
where $a^{[i}b^{j]}=(a^ib^j+a^jb^i)/2-{\bf a}\cdot{\bf b}\;\delta^{ij}/3$.
Thus we are able to decompose the factor $q^i \xi^{\dagger} \sigma^j \eta$
appearing in the
second term on the right-hand-side of Eq.~(\ref{t1red}) into a scalar
component that is identified with the ${}^3P_0$ state, a vector
component identified with the ${}^3P_1$ state, and a
symmetric-traceless tensor component identified with the ${}^3P_2$
state. We wish to emphasize that the procedure of decomposing the
amplitude into seperate angular-momentum configurations, ${}^1S_0$,
${}^3S_1$, ${}^3P_0$, ${}^3P_1$, {\it etc.} can be carried out to any
order in the ${\bf q}^2/m^2_c$ expansion. It is always possible
to decompose the amplitude for the
production of a heavy quark-antiquark pair into pieces of the form
%
%
%
\begin{eqnarray}
{\cal T}(i\rightarrow c\bar{c}({}^1S_0))  &=& A\; \xi^\dagger\eta
\left( 1 + a \frac{{\bf q}^2}{m_c^2} + \cdots \right)
\nonumber\\
{\cal T}(i\rightarrow c\bar{c}({}^3S_1 )) &=& A^i \;\xi^\dagger \sigma^i \eta
\left( 1 + a \frac{{\bf q}^2}{m_c^2} + \cdots \right)
\nonumber\\
{\cal T}(i\rightarrow c\bar{c}({}^1P_1 )) &=& A^i \;\xi^\dagger q^i \eta
\left( 1 + a \frac{{\bf q}^2}{m_c^2} + \cdots \right)
\nonumber\\
{\cal T}(i\rightarrow c\bar{c}({}^3P_0 )) &=&
A \; \xi^\dagger {\bf q} \cdot \mbox{\boldmath{$\sigma$}} \eta
\left( 1 + a \frac{{\bf q}^2}{m_c^2} + \cdots \right)
\nonumber\\
{\cal T}(i\rightarrow c\bar{c}({}^3P_1 )) &=&
A^i \; \xi^\dagger {\bf q} \times \mbox{\boldmath{$\sigma$}}^i \eta
\left( 1 + a \frac{{\bf q}^2}{m_c^2} + \cdots \right)
\nonumber\\
{\cal T}(i\rightarrow c\bar{c}({}^3P_2 )) &=&
A^{ij} \; \xi^\dagger q^{[i}\sigma^{j]} \eta
\left( 1 + a \frac{{\bf q}^2}{m_c^2} + \cdots \right) \nonumber
\end{eqnarray}
where the factors $A$, $A^i$, $A^{ij}$ are independent of
${\bf q}$.

The ${}^3P_0$ contribution can be written as
\begin{equation}
{\cal T}(gg \to c\bar{c}(\underline{8},{}^3 P_0))  =
d^{abc} {4 g^2_s \over (2m_c)^5}  \
\epsilon_\mu(g_1) \epsilon_\nu(g_2)  \;
\left( -2 g^{\mu \nu} m^2_c +
P^\mu g^{\nu}_1 \right)
\xi^{\dagger} {\bf q} \cdot
\mbox{\boldmath$\sigma$}  \; T^c \eta \;  + \cdots  \; .
\label{3p0}
\end{equation}
Squaring and performing the sum-average over
spins and colors, we obtain
\begin{equation}
|\bar{{\cal T}}(gg \to c\bar{c}(\underline{8},{}^3 P_0))|^2 =
{5 g^4_s \over 96 (2m_c)^6 } \; \; \xi^{\dagger} {\bf q} \cdot
\mbox{\boldmath$\sigma$}
T^c \eta \; \;\eta^{\dagger} {\bf q}  \cdot \mbox{\boldmath$\sigma$}
T^c \xi \; \left( 1 + \cdots \right) \; .
\label{sq3p0}
\end{equation}
Inserting Eq.~(\ref{sq3p0})
into Eq.~(\ref{crsc}) we obtain
\begin{equation}
\sigma(gg \to c\bar{c}(\underline{8},{}^3P_0)) =
{5 \pi^3 \alpha^2_s \over 3 \cdot 128 \, m_c^6} \;
\; \int{d^3 {\bf q} \over (2 \pi)^3} \;
\delta(E_f - E_i)
\xi^{\dagger} {\bf q} \cdot
\mbox{\boldmath$\sigma$}
T^c \eta \; \; \eta^{\dagger} {\bf q}  \cdot \mbox{\boldmath$\sigma$}
T^c \xi  \left( 1 + \cdots \right) \; ,
\label{3p0cs}
\end{equation}
which is the result from full QCD. As to the perturbative NRQCD
result, this is
\begin{eqnarray}
\sigma(gg \to c\bar{c}(\underline{8},{}^3P_0)) & = &
\int \frac{d^3 {\bf q}}{(2\pi)^3}
\; \frac{F_8({}^3P_0)}{m_c^4} \;
\langle 0 | {\cal O}_8^{c\bar{c}(q)}({}^3P_0)|0\rangle  + \cdots \nonumber\\
& = & \int \frac{d^3 {\bf q}}{(2\pi)^3}
\; \frac{F_8({}^3P_0)}{m_c^4} \;
\sum_{s\bar{s}m\bar{m}}
{1\over 3}
\langle 0 | \chi^\dagger \left( -\frac{i}{2}
\stackrel{\leftrightarrow}{{\bf D}}\cdot \sigma \right) T^a \psi
| c(s,m,{\bf q})\bar{c}(\bar{s},\bar{m},{-\bf q})\rangle\nonumber\\
&&\qquad
\langle c(s,m,{\bf q})\bar{c}(\bar{s},\bar{m},{-\bf q})|
 \psi^\dagger \left( -\frac{i}{2}
\stackrel{\leftrightarrow}{{\bf D}}\cdot \sigma \right) T^a \chi
| 0 \rangle + \cdots \nonumber\\
&=& \int {d^3 {\bf q} \over (2 \pi)^3} \;
\; \frac{F_8({}^3P_0)}{m_c^4} \;
\frac{1}{3} \;
\xi^{\dagger} {\bf q} \cdot
\mbox{\boldmath$\sigma$}
T^c \eta \;\; \eta^{\dagger} {\bf q}  \cdot \mbox{\boldmath$\sigma$}
T^c \xi \; + \cdots
\label{nrqcd3p0}
\end{eqnarray}
Matching Eqs.~(\ref{3p0cs}) and (\ref{nrqcd3p0}), we obtain
\begin{equation}
F_8({}^3P_0) =
{5 \pi^3 \alpha^2_s \over 32 m_c} \;
\delta ( s - 4m^2_c)
\end{equation}
from which we conclude that
\begin{eqnarray}
\sigma(gg \to \psi + X) &=&
\frac{F_8({}^3P_0)}{m_c^4}\langle 0|  {\cal O}^{\psi}_8 ({}^3P_0) | 0 \rangle
\nonumber\\
&=&{5 \pi^3 \alpha^2_s \over (2m_c)^5} \;
\delta ( s - 4m^2_c) \;
\langle 0|  {\cal O}^{\psi}_8 ({}^3P_0) | 0 \rangle  \; .
\label{psi3p0}
\end{eqnarray}

The ${}^3P_1$ piece in Eq.~(\ref{t1red}) is found to vanish exactly.

Extracting the ${}^3P_2$ contribution from Eq.~(\ref{t1red}), one obtains
\begin{eqnarray}
{\cal T}(gg \to c\bar{c}(\underline{8},{}^3 P_2)) & & =
d^{abc}{ 4 g^2_s \over (2m_c)^4} \;
\epsilon_\mu(g_1) \epsilon_\nu(g_2)
\nonumber \\
\times & & \Bigg\{ 2 m_c \Lambda^{\mu}_i \Lambda^{\nu}_j -
{(g_1 \cdot \Lambda)_i \over m_c} \left(
g^{\mu}_2 \Lambda^{\nu}_j - g^{\nu}_1 \Lambda^{\mu}_j +
g^{\mu \nu} (g_1 \cdot \Lambda)_j \right) \Bigg\}
\nonumber\\
\times & &
\xi^{\dagger} q^{[i} \sigma^{j]} T^c \eta \;  + \cdots  .
\label{3p2}
\end{eqnarray}
Anticipating that the
cross section will involve an integral over $d^3 {\bf q}$,
we note the identity
\begin{eqnarray}
\int {d^3 {\bf q} \over (2 \pi)^3} \; & &
\xi^{\dagger} q^{[i} \sigma^{j]}\; T^c \eta \;\;
\eta^{\dagger} q^{[m} \sigma^{n]}\; T^c\xi =
\nonumber \\
& & {1 \over 5} \left( {\delta^{im} \delta^{jn}
+ \delta^{in}\delta^{jm} \over 2}
- { \delta^{ij}\delta^{mn} \over 3} \right)
\int {d^3 {\bf q} \over (2 \pi)^3} \;
\xi^{\dagger} q^{[l} \sigma^{p]}T^c \eta \;\;
\eta^{\dagger} q^{[l} \sigma^{p]}T^c\xi\; \; ,
\label{tp}
\end{eqnarray}
which, upon squaring  Eq.~(\ref{3p2}) and performing the sum-average over
spins and colors allows us to write
\begin{equation}
\int {d^3 {\bf q} \over (2 \pi)^3} \;
|\bar{{\cal T}}(\underline{8},{}^3 P_2)|^2 = { g^4_s \over 24 (2m_c)^6}
\; \int {d^3 {\bf q} \over (2 \pi)^3} \;
\xi^{\dagger} q^{[i} \sigma^{j]}T^c \eta \;\;
\eta^{\dagger} q^{[i} \sigma^{j]}T^c\xi \; \left( 1 + \cdots \right)  .
\label{sq3p2}
\end{equation}
Substituting Eq.~(\ref{sq3p2}) into
Eq.~(\ref{crsc}), we obtain the full-QCD result
\begin{equation}
\sigma(gg \to c\bar{c}(\underline{8},{}^3P_2)) =
{\pi^3 \alpha^2_s \over 96 m_c^6} \;
\int{d^3 {\bf q} \over (2 \pi)^3} \;
\delta(E_f - E_i) \;
\xi^{\dagger} q^{[i} \sigma^{j]}T^c \eta \;\;
\eta^{\dagger} q^{[i} \sigma^{j]}T^c \xi  \;
\left( 1 + \cdots \right)  .
\label{3p2cs}
\end{equation}
On the other hand, the expression
for $\sigma(gg \to c\bar{c}(\underline{8},{}^3P_2))$
calculated to lowest order in perturbative NRQCD is
\begin{eqnarray}
\sigma(gg \to c\bar{c}(\underline{8},{}^3P_2)) & = &
\int \frac{d^3 {\bf q}}{(2\pi)^3}
\; \frac{F_8({}^3P_2)}{m_c^4} \;
\langle 0 | O_8^{c\bar{c}(q)}({}^3P_2)|0\rangle  + \cdots \nonumber\\
& = & \int \frac{d^3 {\bf q}}{(2\pi)^3}
\; \frac{F_8({}^3P_2)}{m_c^4} \;
\sum_{s\bar{s}m\bar{m}}
\langle 0 | \chi^\dagger \left( -\frac{i}{2}
\stackrel{\leftrightarrow}{D}^{[i}\sigma^{j]} \right) T^a \psi
| c(s,m,{\bf q})\bar{c}(\bar{s},\bar{m},{-\bf q})\rangle\nonumber\\
&&\qquad
\langle c(s,m,{\bf q})\bar{c}(\bar{s},\bar{m},{-\bf q})|
 \psi^\dagger \left( -\frac{i}{2}
\stackrel{\leftrightarrow}{D}^{[i}\sigma^{j]} \right) T^a \chi
| 0 \rangle + \cdots \nonumber\\
& = &
\int {d^3 {\bf q} \over (2 \pi)^3} \;
\frac{F_8({}^3P_2)}{m_c^4} \;
\xi^{\dagger} q^{[i}
\sigma^{j]}
T^c \eta \;\; \eta^{\dagger} q^{[i}  \sigma^{j]}
T^c \xi  + \cdots
\label{nrqcd3p2}
\end{eqnarray}
Matching Eqs.~(\ref{3p2cs}) and (\ref{nrqcd3p2})
we obtain
\begin{equation}
F_8({}^3P_2) =
{\pi^3 \alpha^2_s \over 24 m_c } \;
\delta ( s - 4m^2_c)
\end{equation}
from which we conclude that
\begin{eqnarray}
\sigma(gg \to \psi + X) & = &
\frac{F_8({}^3P_2)}{m_c^4}
\langle 0 |{\cal{O}}^{\psi}_8({}^3P_2) | 0 \rangle
\nonumber\\
& = & {4 \pi^3 \alpha^2_s \over 3 (2m_c)^5} \;
\delta ( s - 4m^2_c) \;
\langle 0 |{\cal{O}}^{\psi}_8({}^3P_2) | 0 \rangle \; .
\label{psi3p2}
\end{eqnarray}

We can now write the
proton-antiproton $\psi$-production cross section by convoluting
the subprocess cross sections
with the
parton distribution functions.
{}From the subprocess cross-section for $q \bar{q} \to \psi +X $, given in
Eq.~(\ref{psics}), we have
\begin{eqnarray}
\sigma(p \bar{p} \to \psi + X) = &&  \int dx_1 f_{q/p}(x_1)
\int dx_2 f_{\bar{q}/p}(x_2)\nonumber\\
&&\qquad {2 \alpha^2_s \pi^3 \over 27 m^3_c} \;
\delta(x_1x_2s-4m^2_c) \; \langle {\cal O}^{\psi}_8({}^3S_1) \rangle
\; \; + \; \; q \to \bar{q} \; ,
\label{inccs1}
\end{eqnarray}
where $s=(P_p + P_{\bar{p}})^2$ is the center-of-mass energy squared
of the colliding proton-antiproton system.
{}From the subprocess cross-sections for
$g g \to \psi +X$, given in
Eq.~(\ref{psi1s0}), Eq.~(\ref{psi3p0}), and Eq.~(\ref{psi3p2}),
we have
\begin{eqnarray}
\sigma(p \bar{p} \to \psi + X) & & = \int dx_1 f_{g/p}(x_1)
\int dx_2 f_{g/p}(x_2)
\nonumber \\
& & \! \! {5 \alpha^2_s \pi^3 \over 12 (2m_c)^3} \;
\delta(x_1x_2s-4m^2_c) \; \left(
\langle {\cal O}^{\psi}_8({}^1S_0) \rangle +
{3 \over m^2_c}
\langle {\cal O}^{\psi}_8({}^3P_0) \rangle +
{4 \over 5 m^2_c}
\langle {\cal O}^{\psi}_8({}^3P_2) \rangle \right) \; .
\nonumber \\
\label{inccs2}
\end{eqnarray}
This expression can be further simplified using the relation
\begin{equation}
\langle {\cal O}^{\psi}_8({}^3P_J) \rangle \approx
(2J+1) \langle {\cal O}^{\psi}_8({}^3P_0) \rangle \; ,
\label{relpwave}
\end{equation}
which holds to within corrections of relative order
$v^2$.


Finally we note a means of checking our results for the subprocess
cross-sections
$gg\to c\bar{c}(\underline{8},{}^{2S+1}L_J)$
given in Eqs.~(\ref{psi1s0}), (\ref{psi3p0}) and
(\ref{psi3p2}).
This check entails first noting
the surprising fact that the amplitudes for
$gg\to c\bar{c}(\underline{8},{}^{2S+1}L_J))$
are proportional to those for
$gg\to c\bar{c}(\underline{1},{}^{2S+1}L_J))$.
Thus, our results for octet production rates
should be
related --- by an overall color-factor and a replacement of
matrix elements -- to results computed
from the same subprocesses for $\eta_c$ and $\chi_c$ production
in the color-singlet-wavefunction
model.  These color-singlet results can be obtained from our
color-octet results
by first performing the replacements
\begin{eqnarray}
\langle {\cal O}^{\psi}_8({}^1S_0) \rangle & \rightarrow &
\frac{15}{8} \langle {\cal O}^{\eta}_1({}^1S_0) \rangle \nonumber\\
\langle {\cal O}^{\psi}_8({}^3P_0) \rangle & \rightarrow &
\frac{15}{8} \langle {\cal O}^{\chi_0}_1({}^3P_0) \rangle \nonumber\\
\langle {\cal O}^{\psi}_8({}^3P_2) \rangle & \rightarrow &
\frac{15}{8} \langle {\cal O}^{\chi_2}_1({}^3P_2) \rangle \; .
\end{eqnarray}
To convert these expressions into the language of the
color-singlet-wavefunction model, one makes the further replacements
\begin{eqnarray}
\langle {\cal O}^{\eta}_1({}^1S_0) \rangle & \rightarrow &
\frac{N_c}{2\pi}|R_s(0)|^2\nonumber\\
\langle {\cal O}^{\chi_{J}}_1({}^3P_J) \rangle & \rightarrow &
\frac{3 (2J+1) N_c}{2\pi}\; |R'_s(0)|^2 \; .
\end{eqnarray}
One observes that, indeed, these operations yield results
which agree with those presented in Ref~\cite{GW}.

\section{Conclusion}


In this paper we have presented a calculation of the hadronic $\psi$
production
cross-section, carried out within the framework of the
NRQCD factorization formalism of Bodwin, Braaten, and Lepage.
We have explicitly shown how to put into practice the ``matching''
procedure, which allows a determination of the short-distance
coefficients appearing in the factorization formula.
By remaining loyal to, and expanding upon, the
matching program briefly described in Ref~\cite{BBL},
we have revealed the true spirit
of the BBL formalism for quarkonium production.
We have also
demonstrated how, operationally, one may systematically include
relativistic corrections.

We have obtained the following results for the leading order
subprocess cross sections due to the color-octet mechanism:
\begin{eqnarray}
\sigma(q\bar{q}\rightarrow \psi) && =
\frac{2 \alpha_s^2 \pi^3}{27 m_c^3}\;\delta(s - 4 m_c^2) \;
\langle 0 | {\cal O}_8^\psi({}^3S_1)| 0 \rangle\nonumber\\
\sigma(gg \rightarrow \psi) && =
\frac{5 \alpha_s^2 \pi^3}{96 m_c^3}\;\delta(s - 4 m_c^2)
\left( \langle 0 | {\cal O}_8^\psi({}^1S_0)| 0 \rangle
+ \frac{3}{m^2_c}
\langle 0 | {\cal O}_8^\psi({}^3P_0)| 0 \rangle
+
\frac{4}{5 m^2_c}
\langle 0 | {\cal O}_8^\psi({}^3P_2)| 0 \rangle \right) \; .
\nonumber
\end{eqnarray}
As a check on these results, we have transformed the above expressions
into color-singlet-wavefunction results for $\eta_c$ and $\chi_c$
production, and have found them to be consistent with previous work.
We wish to point out that our
expression for $\sigma(gg \rightarrow \psi)$ differs, by a factor of 3,
from an analogous expressions obtained in Ref~\cite{tv}.

The formul\ae\ summarized above are of great relevance in the
parametrization of quarkonium production, and will be useful in future
analyses. In particular, these expression, along with the
previously calculated color-singlet result, appear in a
leading order calculation of hadronic $\psi$ production at fixed target
experiments. In fact there are discrepancies between theoretical
predictions and experimental results for $\psi$ production in pion-nucleon
interactions~\cite{bhtv} that will be resolved by including, in the
theoretical predictions, contributions from color-octet
subprocesses~\cite{uspi}.

Also note that the NRQCD matrix elements
$\langle 0 | {\cal O}_8^\psi({}^3S_1)| 0 \rangle $,
$\langle 0 | {\cal O}_8^\psi({}^1S_0)| 0 \rangle$,
$\langle 0 | {\cal O}_8^\psi({}^3P_0)| 0 \rangle$,
and $\langle 0 | {\cal O}_8^\psi({}^3P_2)| 0 \rangle$ that appear in
our calculation, in the linear combinations summarized in the above
formul\ae, will appear in different
linear combinations in a calculation of $\psi$ production
in hadronic colliders~\cite{pcal}.

It must be pointed out that our result for
$\sigma(gg\rightarrow \psi)$ is readily converted into an expression
for the forward photoproduction rate
$\sigma(\gamma g\rightarrow \psi)$, which is the subject of
a work in progress~\cite{us}.


Finally, we mention that while this work was in progress,
there appeared an article by Peter Cho and Adam Leibovich~\cite{pcal}
in which very similar work was presented.  These authors
calculated the leading order $\psi$ production subprocess cross-sections
that we present here, and we agree with their results.
We must point out however that the approach to matching
in~\cite{pcal} is quite different from ours, and that
the two articles are complementary.

\acknowledgements


It is our pleasure to thank Peter Lepage for
helpful discussions. It is also our pleasure to thank Eric Braaten for
his generosity, helpful discussions, and for providing us with the
reduction formul\ae\ in Eqs.~(\ref{red1}) and (\ref{red2}).
We thank Peter Cho and Adam Leibovich for pointing out
some typographical errors in the original manuscript, and
Patrick Labelle for making
some helpful comments.
The work of I.M. was
supported by the Robert A. Welch Foundation, by NSF Grant PHY 9009850,
and by NSERC of Canada.
The work of S.F. was supported in part by the U.S.~Department of Energy under
Grant No.~DE-FG02-95ER40896, in part by the University of Wisconsin Research
Committee with funds granted by the Wisconsin Alumni Research Foundation, and
in part by the National Science Foundation under Grant No.~PHY94-07194.

\vfill\eject





\vbox{

\begin{table}[h]
\caption{The $v$-scaling of the NRQCD color-octet matrix elements,
relative to $S$-wave baseline.}
\label{vmes}
\smallskip
\begin{tabular}{cccc}
 & $\langle 0| {\cal O}_8^H ({}^{(2S+1)}L_J) | 0 \rangle$
& Scaling
\\ \hline
   & $\langle 0| {\cal O}^{\psi}_1({}^3S_1) |0 \rangle $  & $v^{0}$
\\ & $\langle 0| {\cal O}^{\psi}_8({}^3S_1) |0 \rangle $  & $v^{4}$
\\ & $\langle 0| {\cal O}^{\psi}_8({}^1S_0) |0 \rangle $  & $v^{4}$
\\ & $\langle 0| {\cal O}^{\psi}_8({}^3P_J) |0 \rangle /m_c^2 $  & $v^{4}$
\end{tabular}
\end{table}

\begin{table}[h]
\caption{The scaling with $\alpha_s(m_c)$ of the short-distance
coefficients}
\label{sdcs}
\smallskip
\begin{tabular}{cccc}
 & $ i + j \to c\bar{c}(n) + X $ & Scaling
\\ \hline
 & $gg \to c\bar{c}(\underline{1},{}^3S_1)+g$           & $\alpha_s(m_c)^3$
\\
 & $q\bar{q} \to c\bar{c}(\underline{8},{}^3S_1)$       & $\alpha_s(m_c)^2$
\\
 & $gg \to c\bar{c}(\underline{8},{}^1S_0,{}^3P_{0,2})$ & $\alpha_s(m_c)^2$
\\
\end{tabular}
\end{table}

}

\begin{figure}
\begin{center}
\mbox{\epsfig{file=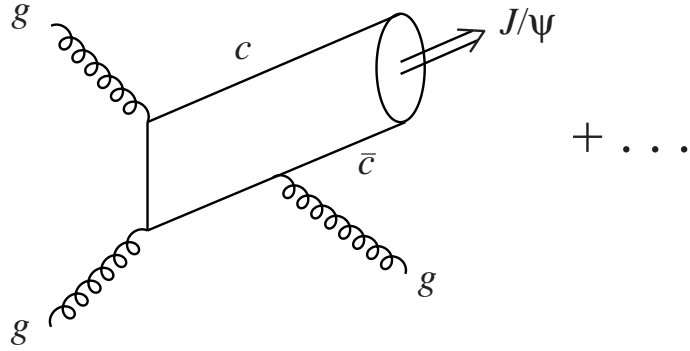,height=5.0cm}}
\end{center}
\caption{Feynman diagrams for leading-order color-singlet
hadronic production of charmonium.}
\label{locsfig}
\end{figure}
\begin{figure}
\begin{center}
\mbox{\epsfig{file=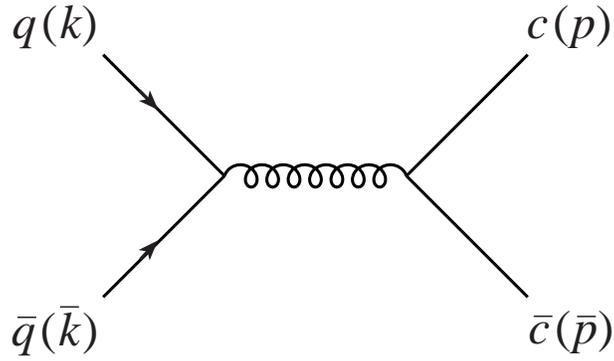,height=5.0cm}}
\end{center}
\caption{Feynman diagram for the color-octet
subprocess $q\bar{q} \to c\bar{c}(\underline{8},{}^3S_1)$.}
\label{qqbarfig}
\end{figure}

\pagebreak

\begin{figure}
\begin{center}
\mbox{\epsfig{file=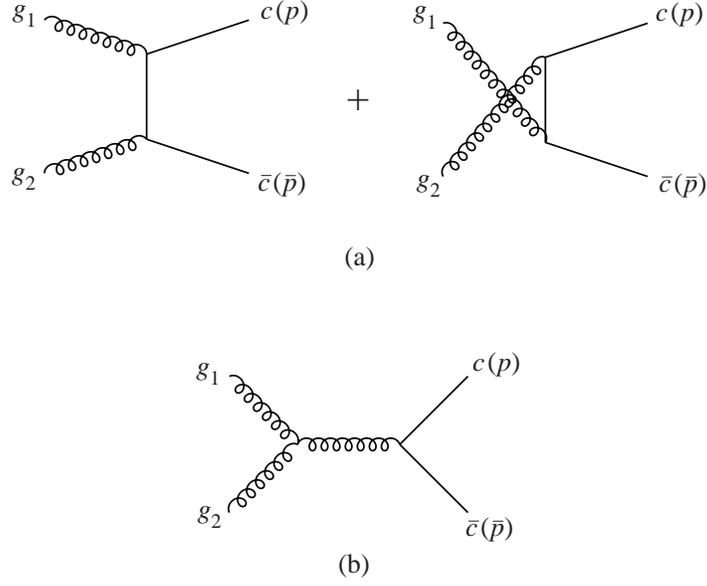,height=8.0cm}}
\end{center}
\caption{Color-octet subprocesses
$gg \to c\bar{c}(\underline{8},{}^1S_0,{}^3S_1,{}^3P_{0,2})$.
Figure 3a: Gluon fusion Feynman diagram,
which produces
$c\bar{c}(\underline{8}, {}^1S_0, {}^3S_1, {}^3P_{0,2})$.
Figure 3b: Feynman diagram for $gg\to g \to c\bar{c}
(\underline{8},{}^3S_1)$.  At leading order in the
non-relativistic expansion, the amplitude for
$c\bar{c}(\underline{8},{}^3S_1)$ production in
Figure 3a cancels against that in Figure 3b.}
\label{ggfig}
\end{figure}

\end{document}